\documentclass[onecolumn]{aastex62}

\usepackage{lineno}
\usepackage{longtable}

\graphicspath{./}

\begin{document}

\title{A catalog of new slowly pulsating B-type stars}

\correspondingauthor{Xiang-dong Shi}
\email{sxd@ynao.ac.cn}

\author{Xiang-dong Shi}
\affiliation{Yunnan Observatories, Chinese Academy of Sciences(CAS), P.O. Box 110, Kunming 650216, P.R. China}
\affiliation{Department of Astronomy, Key Laboratory of Astroparticle Physics of Yunnan Province, Yunnan University, Kunming 650091, P.R. China}
\affiliation{University of Chinese Academy of Sciences, No.1 Yanqihu East Rd, Huairou District, Beijing 101408, P.R. China}
\affiliation{Key Laboratory of the Structure and Evolution of Celestial Objects, CAS, Kunming 650216, P.R. China}

\author{Sheng-bang Qian}
\affiliation{Yunnan Observatories, Chinese Academy of Sciences(CAS), P.O. Box 110, Kunming 650216, P.R. China}
\affiliation{Department of Astronomy, Key Laboratory of Astroparticle Physics of Yunnan Province, Yunnan University, Kunming 650091, P.R. China}
\affiliation{University of Chinese Academy of Sciences, No.1 Yanqihu East Rd, Huairou District, Beijing 101408, P.R. China}
\affiliation{Key Laboratory of the Structure and Evolution of Celestial Objects, CAS, Kunming 650216, P.R. China}

\author{Li-ying Zhu}
\affiliation{Yunnan Observatories, Chinese Academy of Sciences(CAS), P.O. Box 110, Kunming 650216, P.R. China}
\affiliation{University of Chinese Academy of Sciences, No.1 Yanqihu East Rd, Huairou District, Beijing 101408, P.R. China}
\affiliation{Key Laboratory of the Structure and Evolution of Celestial Objects, CAS, Kunming 650216, P.R. China}

\author{Lin-jia Li}
\affiliation{Yunnan Observatories, Chinese Academy of Sciences(CAS), P.O. Box 110, Kunming 650216, P.R. China}
\affiliation{University of Chinese Academy of Sciences, No.1 Yanqihu East Rd, Huairou District, Beijing 101408, P.R. China}
\affiliation{Key Laboratory of the Structure and Evolution of Celestial Objects, CAS, Kunming 650216, P.R. China}

\begin{abstract}
This paper reports the discovery of new slowly pulsating B-type stars. Based on the photometric, spectral, and astrometric data of TESS, LAMOST, and Gaia surveys, we have found 286 new slowly pulsating B-type stars (SPB stars) and 21 candidates. Among these, 20 are Be stars or candidates with emission line profiles. It is shown that these SPB stars have luminosities between 40 and 2850 $L_{\odot}$ and effective temperatures ranging from 10000K to 21000K. Their pulsation periods are from 0.14 to 6.5 days with amplitude ranges of 0.2-20 mmag in TESS band. It is indicated that these targets follow the distribution of the SPB stars in the period-luminosity (P-L) and the period-temperature (P-T) diagrams. Their positions on the H-R diagram reveal that most of these pulsators are distributed in the instability region of SPB stars, in the main-sequence evolutionary stage, and with mass ranges of 2.5-7 $M_{\odot}$. However, there are some targets beyond the red edge of the theoretical instability region, which should be caused by the rapid rotation reducing the measured effective temperature. The discovery of these new SPB stars increases the total number by over 60\%, which are significant samples for further investigating the structure and evolution of intermediate-mass and even massive stars by asteroseismology.
\end{abstract}

\keywords{stars: intermediate-mass -
          stars: pulsating }

\section{Introduction} \label{sec:intro}

Generally, the recognized upper main-sequence pulsating stars on the H-R diagram include slowly pulsating B-type stars (SPB stars) and $\beta$ Cephei pulsating variable stars (BCEP stars). Among them, the SPB stars oscillate in nonradial, high-order, and low-degree g-mode, with period ranges of 0.5-3 days and a spectral type of B-type (e.g., \citet{2010aste.book.....A}). The $\kappa$ mechanism, driven by the ionization zone of iron-group elements, excites the pulsation of SPB stars (e.g., \citet{2016ApJ...823..130M}). The g-mode in SPB stars provides an opportunity to study the interior structure information, that include the size of convective core, the chemical mixing in stellar interiors, the overshooting, rotation, magnetic field of convective core, the mass fraction of hydrogen in the stellar center, and so on (e.g., \citet{2015A&A...580A..27M, 2016ApJ...823..130M, 2018ApJ...867...47W, 2019ApJ...881...86W, 2022MNRAS.512L..16L, 2022ApJ...940...49P, 2021NatAs...5..715P}). Therefore, SPB stars are extremely critical for solving the issues related to internal element-mixing, internal angular momentum transport, stellar wind mass loss, and magnetic activity in the structure and evolution of intermediate-mass and even massive stars.

In recent years, more and more SPB stars have been observed by ground- and space-based surveys (e.g. \citet{2009A&A...506..471D, 1999A&A...343..872A, 2001A&A...379..905M, 2002A&A...393..965D, 2007A&A...463..243D, 2011MNRAS.413.2403B}, etc.). On 29 May 2023, 214 SPB stars were listed in the VSX catalog (the international variable star index, \citet{2006SASS...25...47W}). So far, the number of detected SPB stars may be around 400 to 500, including some targets that have been published but are not included in the VSX catalog, for example, some of the 308 SPB stars detected by \citet{2020MNRAS.493.5871B}. Due to the small sample size (e.g. more than 15,000 $\delta$ Sct stars in VSX catalog), many characteristics of SPB stars may not be clear. Therefore, it is important to discover more samples.


Thanks to the large-scale ongoing surveys such as Transiting Exoplanet Survey Satellite (TESS, \citet{2015JATIS...1a4003R}), Large Sky Area Multi-Object Fiber Spectroscopic Telescope (LAMOST, \citet{1996ApOpt..35.5155W, 2012RAA....12.1197C}), Gaia Satellite \citep{2016A&A...595A...1G, 2018A&A...616A...1G, 2021A&A...649A...1G}, the massive and high-precision data allow us to systematically study all kinds of variables. In particular, due to the observational characteristics of SPB stars, such as low-amplitude and long-period, the high-precision and continuous time-series photometry data of space telescope have incomparable advantages for their discovery and research (e.g., \citet{2019ApJ...872L...9P}). By using TESS, LAMOST, and Gaia data, this paper reports the discovery of new SPB stars, and analyze the preliminary pulsation properties of these objects.

\begin{table*}[]
\begin{center}
\caption{The Catalog of new SPB stars observed by TESS, LAMOST, and GAIA (the First 20 Lines of the Whole Table)}
\label{tab:1}
\begin{tabular}{lllllllllllllll}
 \hline
TESS ID     & $\pi$    & V       & $Teff$  & Flag   & $log(L_/L_{\odot})$   & Comments & Period     & Amplitude \\
            & ($mas$)  & ($Mag$) & ($K$)   &        &                       &          & ($days$)   & ($mmag$)  \\

\hline

1940542   & 1.00 & 11.77 & 10830(93)  & 2       & 1.57 & SPB    & 0.3298385(1)  & 2.02(4)   \\
1951280   & 0.72 & 12.53 & 12596(296) & 2       & 1.87 & SPB    & 0.7566529(5)  & 4.11(6)   \\
2105758   & 1.01 & 10.62 & 14232(267) & 1       & 2.36 & SPB    & 0.6362165(1)  & 12.67(2)  \\
2342458   & 1.16 & 9.72  & 14466(281) & 1       & 2.59 & SPB    & 0.2100363(1)  & 1.30(1)   \\
2438355   & 0.83 & 10.94 & 11536(122) & 1       & 2.25 & SPB    & 0.3347354(1)  & 0.72(3)   \\
2845400   & 0.70 & 12.28 & 12720(196) & 3       & 2.02 & SPB(C) & 1.852164(7)   & 4.46(5)   \\
3023821   & 0.49 & 11.36 & 17705(373) & 1       & 3.33 & SPB    & 1.273005(2)   & 5.21(4)   \\
3117265   & 1.28 & 10.13 & 12717(161) & 1       & 2.26 & SPB    & 0.7876780(2)  & 1.91(2)   \\
3315867   & 0.45 & 11.84 & 19516(398) & 1(Be?)  & 3.14 & SPB    & 0.5896331(2)  & 0.92(5)   \\
9049366   & 1.16 & 11.19 & 14358(204) & 1       & 2.63 & SPB    & 1.973178(3)   & 3.13(2)   \\
9131633   & 1.31 & 11.22 & 13037(143) & 1       & 2.14 & SPB    & 6.5564(2)     & 12.92(2)  \\
9372100   & 1.34 & 9.60  & 20975(455) & 1       & 3.24 & SPB    & 1.660494(1)   & 1.44(1)   \\
9953044   & 0.58 & 12.44 & 15788(256) & 1       & 2.68 & SPB    & 0.9207100(9)  & 18.72(5)  \\
10069057  & 0.75 & 10.70 & 12115(246) & 1       & 2.24 & SPB    & 5.658173(2)   & 0.37(1)   \\
12312337  & 0.84 & 11.75 & 12780(145) & 1       & 2.06 & SPB    & 1.0663657(1)  & 8.16(2)   \\
13792065  & 1.03 & 11.39 & 14388      & 3(Be)   & 2.91 & SPB    & 0.4926890(1)  & 1.91(2)   \\
21036413  & 0.57 & 10.18 & 12920(184) & 1       & 3.01 & SPB    & 0.4941412(1)  & 0.62(2)   \\
21390403  & 0.56 & 11.00 & 14692(242) & 1       & 2.83 & SPB(C) & 0.6498741(2)  & 2.57(3)   \\
24798057  & 1.97 & 9.00  & 11787(96)  & 1       & 2.04 & SPB(C) & 0.3042837(1)  & 0.71(1)   \\
27252870  & 0.93 & 11.25 & 15056(301) & 1       & 2.60 & SPB    & 0.6065840(1)  & 1.75(2)   \\

\hline
\end{tabular}
\end{center}
\tablecomments{Note. This table is available in its entirety in machine-readable form. The numbers in the parentheses represent the errors of the data. The objects that may be contaminated by nearby stars in the TESS photometry apertures are denoted by the capital letters C in the parenthesis of Column (7). These objects with the $H_{\alpha}$ emission line profile detected from LAMOST are marked as Be in the parentheses of Column (5). The corresponding source flag of effective temperature Teff is given in column 5 as follows: 1 - Gaia ESP-HS; 2 - LAMOST$\_$Xiang; 3 - LAMOST$\_$Guo. \\}
\end{table*}

\begin{table*}[]
\begin{center}
\caption{The candidates of SPB stars observed by TESS, LAMOST, and GAIA}
\label{tab:2}
\begin{tabular}{lllllllllllllll}
 \hline
TESS ID     & $\pi$    & V       & $Teff$  & Flag   & $log(L_/L_{\odot})$  & Comments \\
            & ($mas$)  & ($Mag$) & ($K$)   &        &                      &          \\

\hline

2105067   & 1.08 & 9.87  & 14052(270)  & 1 & 2.58  & SPB+EB        \\
26465775  & 0.54 & 11.65 & 13832(286)  & 1 & 2.32  & SPB+ROT(C)    \\
91782061  & 0.76 & 11.25 & 14056(192)  & 1 & 2.30  & SPB+ROT       \\
104982501 & 1.03 & 11.51 & 11824(93)   & 1 & 2.25  & SPB+ROT       \\
115634558 & 0.84 & 11.56 & 12035(123)  & 1 & 2.22  & SPB+ROT       \\
117325099 & 1.07 & 10.23 & 13760(220)  & 1 & 2.76  & SPB?          \\
117401515 & 0.66 & 12.92 & 17247(335)  & 1 & 2.45  & SPB+ROT       \\
122882943 & 0.91 & 9.73  & 14156(164)  & 1 & 2.78  & SPB+EB        \\
187425112 & 0.73 & 10.71 & 13989(216)  & 1 & 2.64  & SPB?          \\
220195114 & 0.97 & 10.6  & 15697(340)  & 1 & 2.58  & SPB?          \\
234822060 & 0.62 & 11.43 & 19404(329)  & 1 & 3.30  & SPB?          \\
234930473 & 1.42 & 11.13 & 16761(602)  & 1 & 2.18  & SPB?          \\
239786273 & 0.98 & 10.48 & 15559(272)  & 1 & 2.58  & SPB+ROT(C)    \\
241088448 & 0.58 & 11.76 & 11368(90)   & 1 & 1.74  & SPB?          \\
281808654 & 0.74 & 11.60 & 15936(666)  & 1(Be) & 2.35  & SPB?      \\
302677757 & 2.48 & 9.61  & 11282(72)   & 1 & 3.04  & SPB?(C)       \\
417310388 & 1.76 & 9.52  & 12881(153)  & 1 & 1.88  & SPB?          \\
431891272 & 1.00 & 9.835 & 17790(395)  & 1 & 2.94  & SPB?          \\
440492526 & 1.77 & 9.52  & 10908(94)   & 1 & 2.50  & SPB?          \\
450222476 & 0.59 & 12.07 & 17288(202)  & 1 & 2.26  & SPB+EB        \\
470536821 & 0.84 & 10.82 & 14372(352)  & 1 & 2.53  & SPB?          \\

\hline
\end{tabular}
\end{center}
\tablecomments{Note. Same as in Table \ref{tab:1}. \\}
\end{table*}

\section{New slowly pulsating B-type stars observed by TESS, LAMOST, and GAIA} \label{}

LAMOST with a field of view of 5 square degrees can obtain 4000 spectra in a single exposure, equipped with a low-resolution spectrograph (blue arm 3700-5900 $\AA$ and red arm 5700-9000 $\AA$) and a medium-resolution spectrograph (blue arm 4950-5350 $\AA$ and red arm 6300-6800 $\AA$). It has already observed massive stellar spectra, including many pulsating stars (e.g., \citet{2018MNRAS.475..478Q, 2019RAA....19....1Q}) and binary systems (e.g., \citet{2017RAA....17...87Q, 2018ApJS..235....5Q, 2020RAA....20..163Q, 2019ApJS..244...43Z}).
LAMOST does not officially provide stellar atmospheric parameters for early-type stars, but some researchers have used their own methods to obtain some results from LAMOST spectra. \citet{2022A&A...662A..66X} obtained the stellar atmospheric parameters for about 330,000 OBA stars, using the full spectral fitting tool PAYNE on the low-resolution LAMOST spectra.
\citet{2021ApJS..257...54G} gave the stellar atmospheric parameters for about 20,000 OB stars, using machine learning techniques on the low- and medium-resolution LAMOST spectra.


Gaia Satellite was launched on 19 December 2013 by the European Space Agency (ESA) and has obtained very high-precision astrometry data for nearly 2 billion stars \citep{2016A&A...595A...1G, 2018A&A...616A...1G, 2021A&A...649A...1G}. Its parallax value is an essential parameter for estimating stellar luminosity, which is an independent method for the study of variable stars (e.g., \citet{2019MNRAS.485.2380M}) or the reliability check of research results (e.g., \citet{2021PASP..133e4201S}). The stellar atmospheric parameters are also derived in Gaia DR3 from the Blue and Red Photometer low-resolution spectral data and the Radial Velocity Spectrometer spectra by different methods, including the General Stellar Parametrizer from Photometry (GSP-Phot) and the Extended Stellar Parametrizer for Hot Stars (ESP-HS), etc.

TESS is an MIT-led NASA mission to detect the transiting exoplanets in all-sky. The all-sky is divided into 26 sectors. For every sector, TESS downlink a full-frame image (FFI) every 30 minutes and downlink 2-min short cadence data for about 20,000 targets. In addition to the primary TESS mission, its high-precision and massive photometric data are also very useful for detecting and studying bright variables (e.g., \citet{2020MNRAS.493.5871B, 2022ApJS..259..50S}).

It can be seen from the characteristics of SPB stars that there is almost no pulsation frequency exceeding 10 cycles $d^{-1}$, even in the cases where high rotation shifts g modes to a higher frequency domain (e.g., \citet{2010aste.book.....A, 2007CoAst.151...48M, 2023ApJS..265...33S}, etc.). In addition to the 2-min short cadence data, the FFI data with a 30-min cadence can also well satisfy the requirements of the study of SPB stars, because their Nyquist frequency is 24 cycles $d^{-1}$. Therefore, this work continues the previous work (Paper I, \citet{2023ApJS..265...33S}) and expands the samples to the FFI data processed by the Science Processing Operations Center (SPOC).

The light curves are downloaded from the Mikulski Space Telescope Archive (MAST) database and processed using the steps described by \citet{2021AJ....161...46S, 2021MNRAS.505.6166S}. The steps include converting the fluxes to magnitudes and subtracting the average of magnitudes, which will result in light curves similar to the differential photometry. There are two kinds of TESS light curves provided by SPOC \citep{2010ApJ...713L..87J}: One is the Simple Aperture Photometry (SAP) flux, which represents the flux sum of the calibrated pixels within the TESS optimal photometric aperture; the other is the Pre-search Data Conditioned Simple Aperture Photometry (PDCSAP), which represents that the long-term trends have been removed using so-called Co-trending Basis Vectors, i.e. the instrumental variations and excessive scattered light have been removed. Here, we choose to use PDCSAP data.

Through a simple program aided visual classification, 307 SPB stars or candidates are detected. The criteria of visual classification include, but are not limited to, excluding some classic light curves (such as eclipsing binary and rotational modulation stars), determining their position in the H-R diagram, and where there are significant variations in the light curves and the Fourier spectral features of SPB stars that will be described in the next section. The information of these 307 objects is listed in Tables \ref{tab:1} and \ref{tab:2}.

In Tables \ref{tab:1} and \ref{tab:2}, the parallax $\pi$ and the visual magnitude $m_{V}$ of these targets are from Gaia DR3 and the TESS Input Catalogue (TIC, \citet{2018AJ....156..102S}) or SIMBAD \citep{2000A&AS..143....9W}, respectively. Meanwhile, SIMBAD and Gaia are also used to determine if nearby stars may be contaminating the TESS photometry apertures. According to 21 arcsec per pixel of resolution for TESS, a target will be considered as potentially contaminated, if there is a star no more than 6 magnitudes fainter than our target within a 1 arcmin field of view centred on our target. It is denoted by a capital C in parenthesis at Column (8) of Table \ref{tab:1} and Column (7) of Table \ref{tab:2}.

\begin{figure*}\centering \vbox to1.5in{\rule{0pt}{5.0in}}
\includegraphics{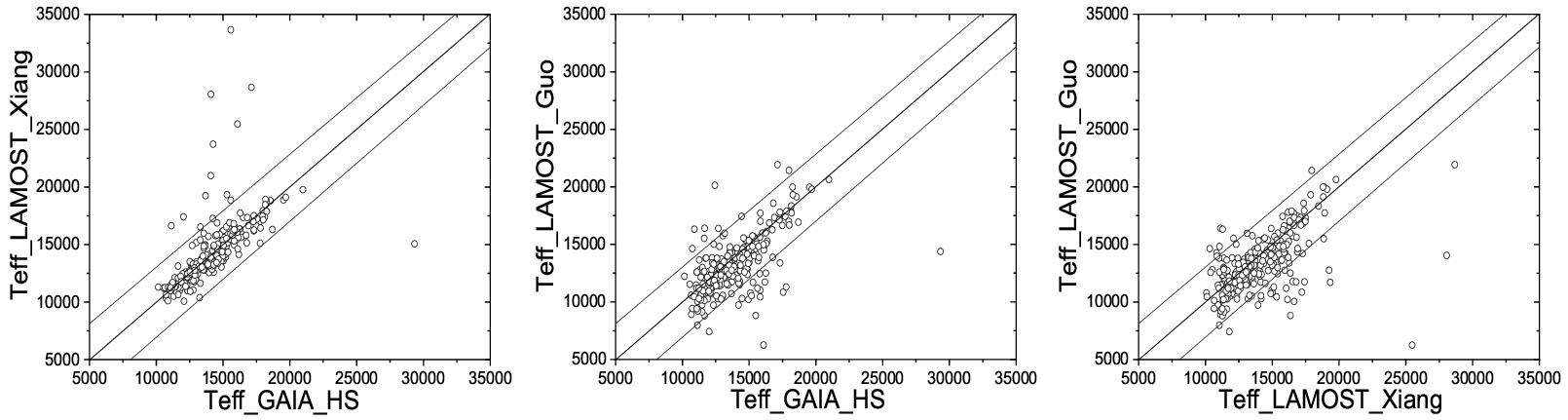}
\caption{The comparison between the temperature of Gaia ESP-HS, that of LAMOST derived by \citet{2022A&A...662A..66X} (LAMOST$\_$Xiang) and by \citet{2021ApJS..257...54G} (LAMOST$\_$Guo). Except for a few targets, most give consistent results.}
\label{fig:T-T}
\end{figure*}

In Figure \ref{fig:T-T}, we compared the temperature of these objects from Gaia ESP-HS and LAMOST derived by \citet{2022A&A...662A..66X} (LAMOST$\_$Xiang) and by \citet{2021ApJS..257...54G} (LAMOST$\_$Guo), and concluded that these temperatures are consistent and reliable in most cases. Meanwhile, based on the principle that the values with larger deviations among the three sources are unreliable for the same target, it can be seen from this figure that LAMOST$\_$ Guo gave a significantly lower temperature (less than 10000K) for some targets, LAMOST$\_$ Xiang gave an obviously high temperature for some targets, and Gaia ESP-HS gave an obviously high temperature for one target. Therefore, we gave the first, second, and third priority corresponding to Gaia ESP-HS, LAMOST$\_$Xiang, and LAMOST$\_$Guo, respectively. Their effective temperatures $Teff$ and corresponding source flags are given in Columns (4) and (5) of Tables \ref{tab:1} and \ref{tab:2}. The parenthesis of Column (4) in Table \ref{tab:1} and \ref{tab:2} list the uncertainty of effective temperature provided by Gaia ESP-HS and LAMOST$\_$Xiang. According to \citet{2021ApJS..257...54G}, the effective temperature has average uncertainties of 1642 K for low-resolution spectra and 2185 K for medium-resolution spectra, respectively. Because low-resolution spectra cover a larger range of wavelengths than medium-resolution spectra, the effective temperatures from low-resolution spectra have less uncertain than those from medium-resolution spectra. But we also try to compare other atmospheric parameters for these three sources and failed to obtain a similar consistent relationship.

\begin{figure*}\centering \vbox to2.5in{\rule{0pt}{5.0in}}
\includegraphics{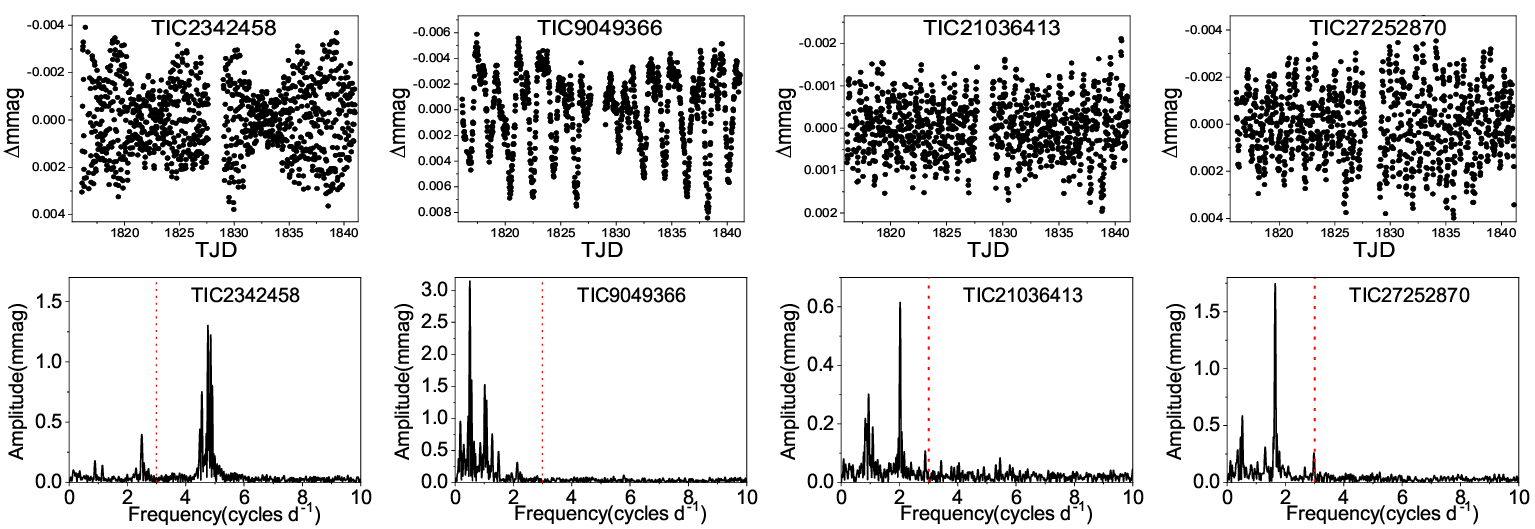}
\caption{Example light curves and Fourier spectra for some SPB stars in the theoretical instability region. The rough boundary to classify high- and low- frequencies is shown as the red dotted line.}
\label{fig:1}
\end{figure*}

\begin{figure*}\centering \vbox to2.5in{\rule{0pt}{5.0in}}
\includegraphics{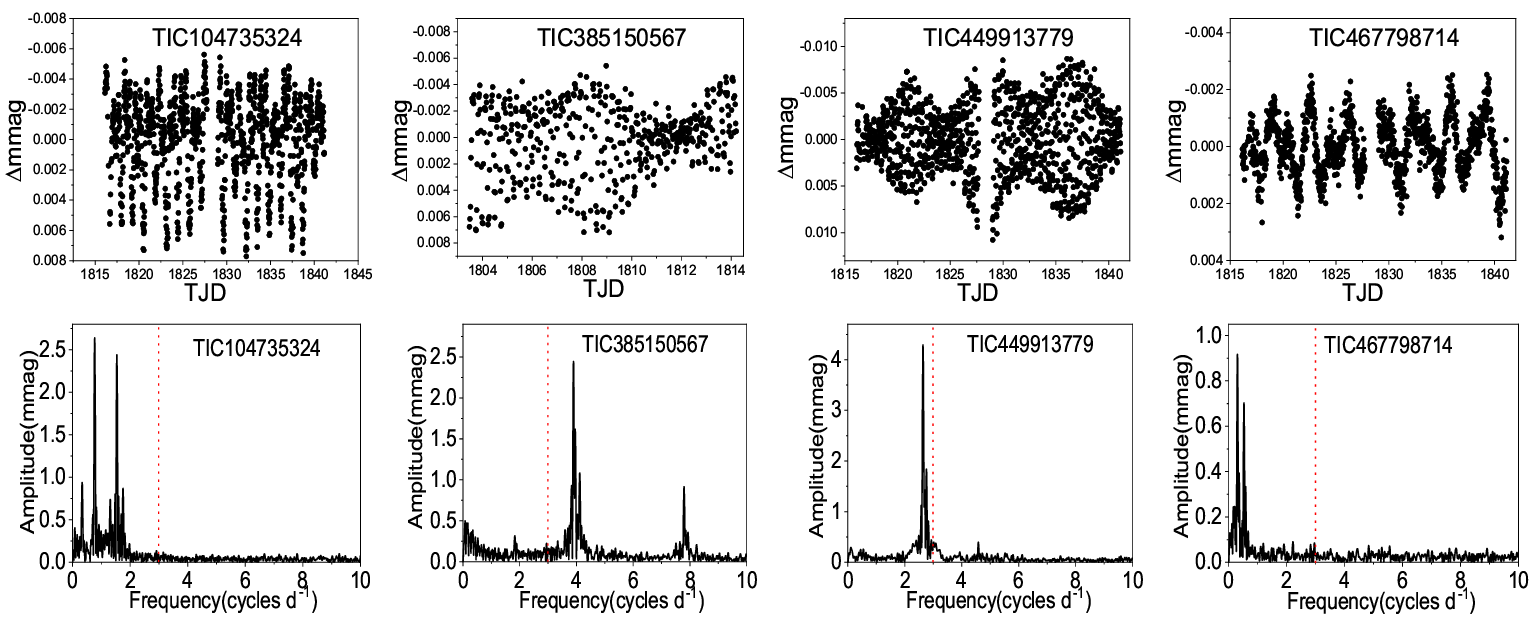}
\caption{Same as in Figure \ref{fig:1} but for stars outside the theoretical SPB instability strip.}
\label{fig:2}
\end{figure*}

\begin{figure*}\centering \vbox to4.0in{\rule{0pt}{5.0in}}
\includegraphics{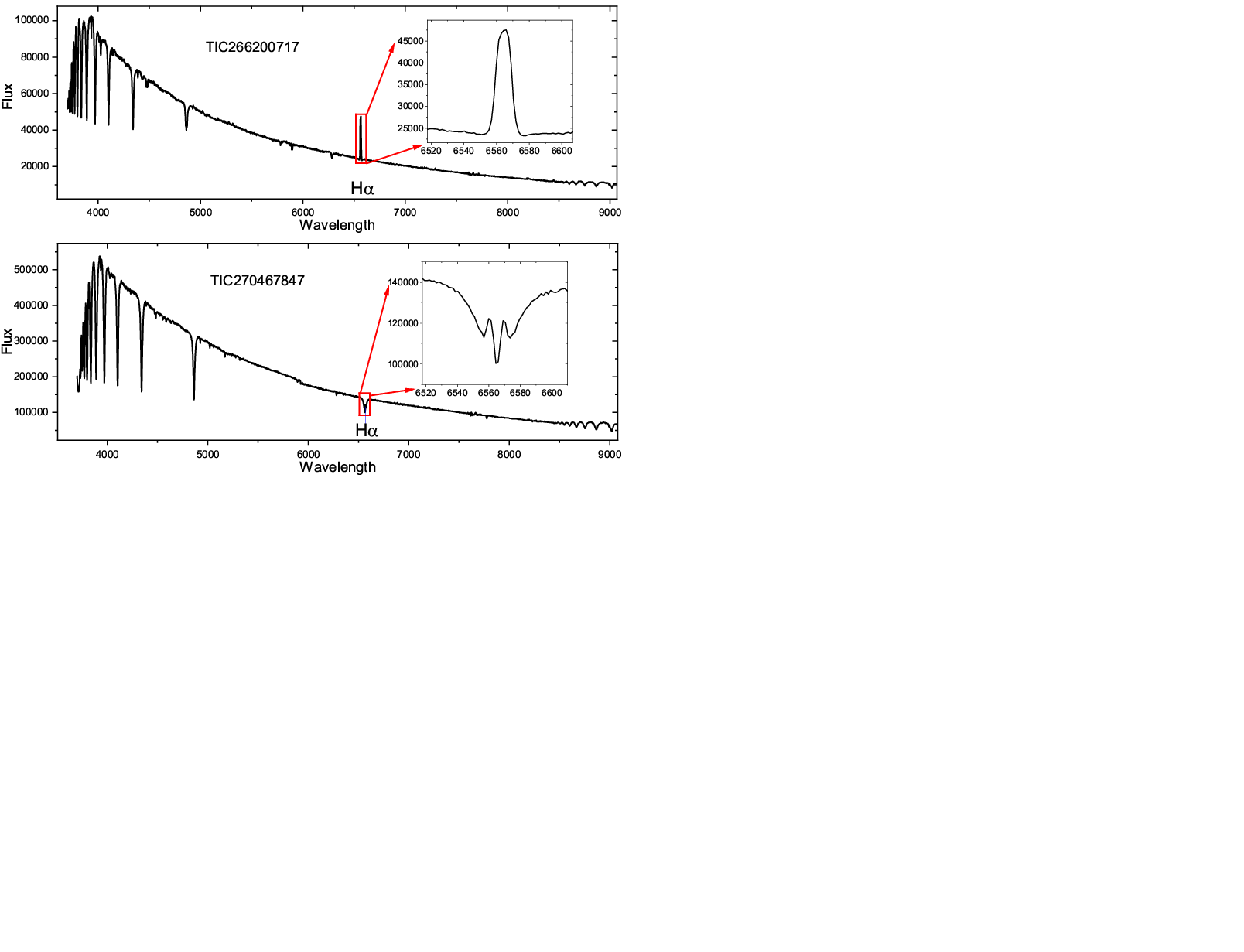}
\caption{Example low-resolution spectra of Be star TIC266200717 and candidate TIC270467847 from LAMOST.}
\label{fig:Spectra}
\end{figure*}

\begin{figure*}\centering \vbox to4.0in{\rule{0pt}{5.0in}}
\includegraphics{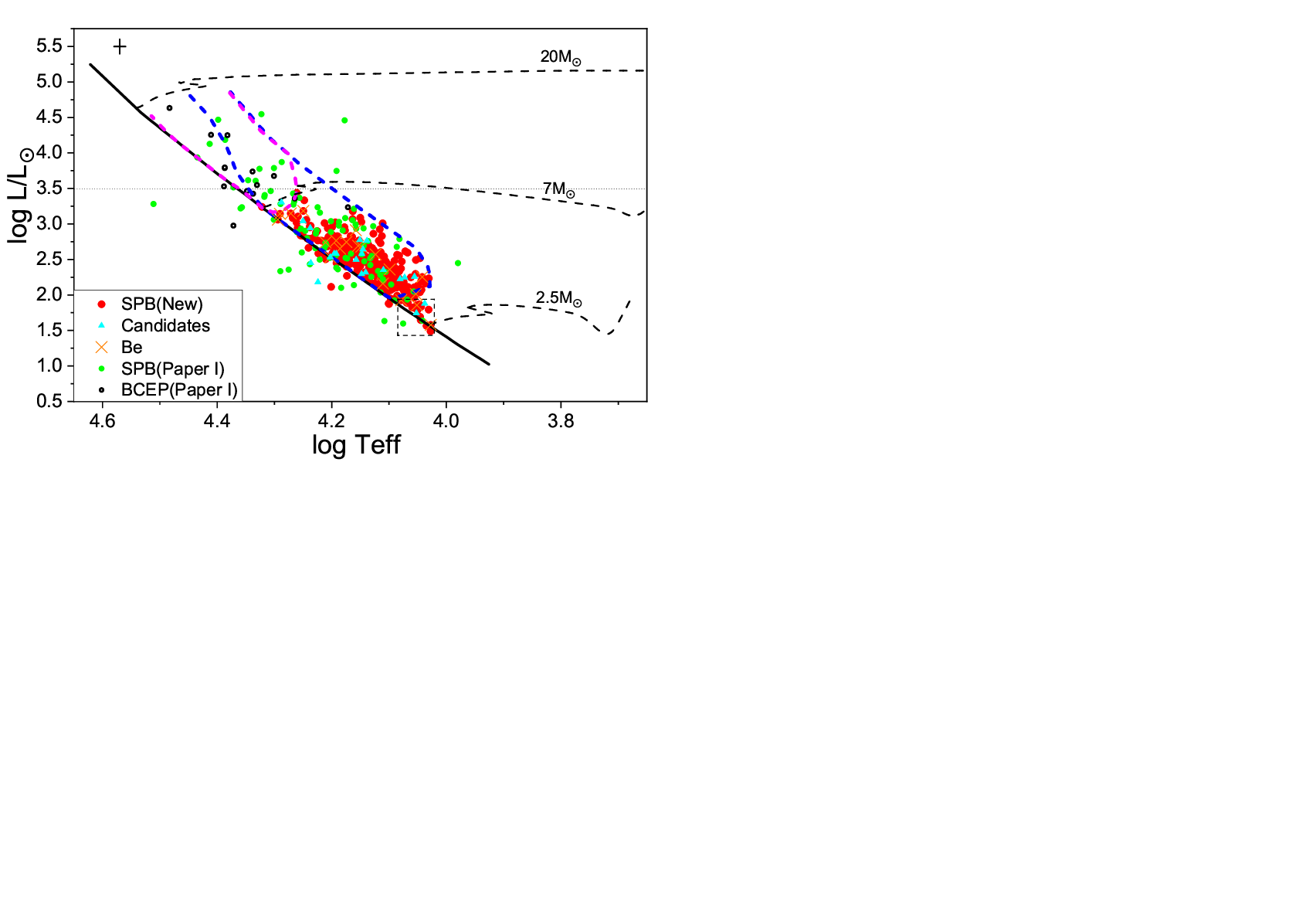}
\caption{The H-R diagram of these new SPB stars. The red solid circles and the cyan solid triangles are these new SPB stars and candidates, respectively. The orange crosses refer to those objects with the $H_{\alpha}$ emission line profile. Meanwhile, those SPB and BCEP stars published in Paper I are also shown as green solid circles and black open circles, respectively. The theoretical zero-age main sequence and evolutionary paths for the various masses with Z=0.02 are represented by the black solid and dotted lines. The theoretical instability region of SPB and BCEP stars for Z = 0.02 and spherical harmonic degree $l \leq$ 3 from \citet{2007CoAst.151...48M} are represented by the blue and magenta dotted lines, respectively. The black cross in the upper left corner represents a typical error box. There are some targets beyond the red edge of theoretical instability region, i.e. these targets in the black dashed box.}
\label{fig:L-T}
\end{figure*}

\begin{figure*}\centering \vbox to2.5in{\rule{0pt}{5.0in}}
\includegraphics{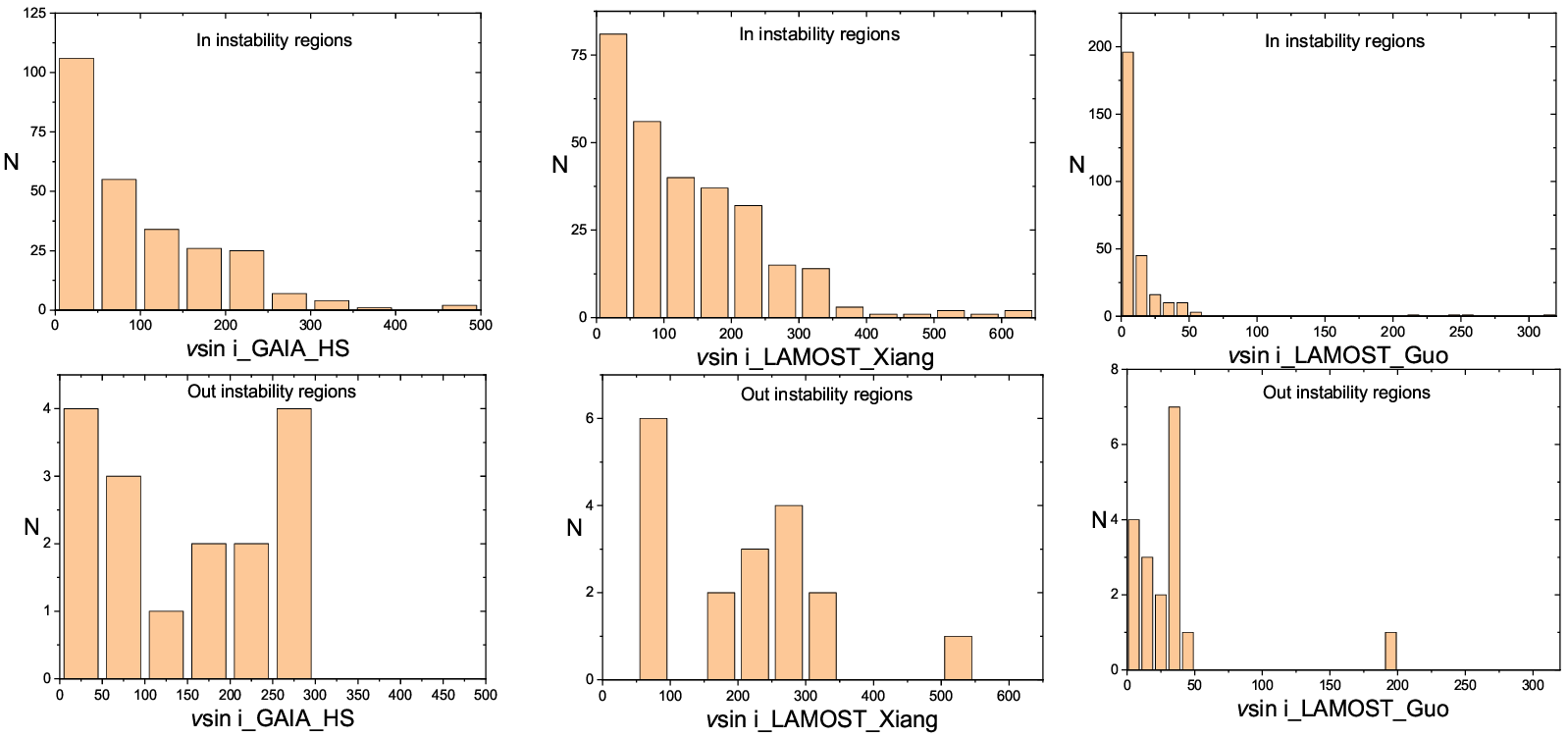}
\caption{Distribution of the projected rotation velocity from the three sources for SPB stars inside and outside the theoretical instability region.}
\label{fig:vsini}
\end{figure*}

\begin{figure*}\centering \vbox to4.0in{\rule{0pt}{5.0in}}
\includegraphics{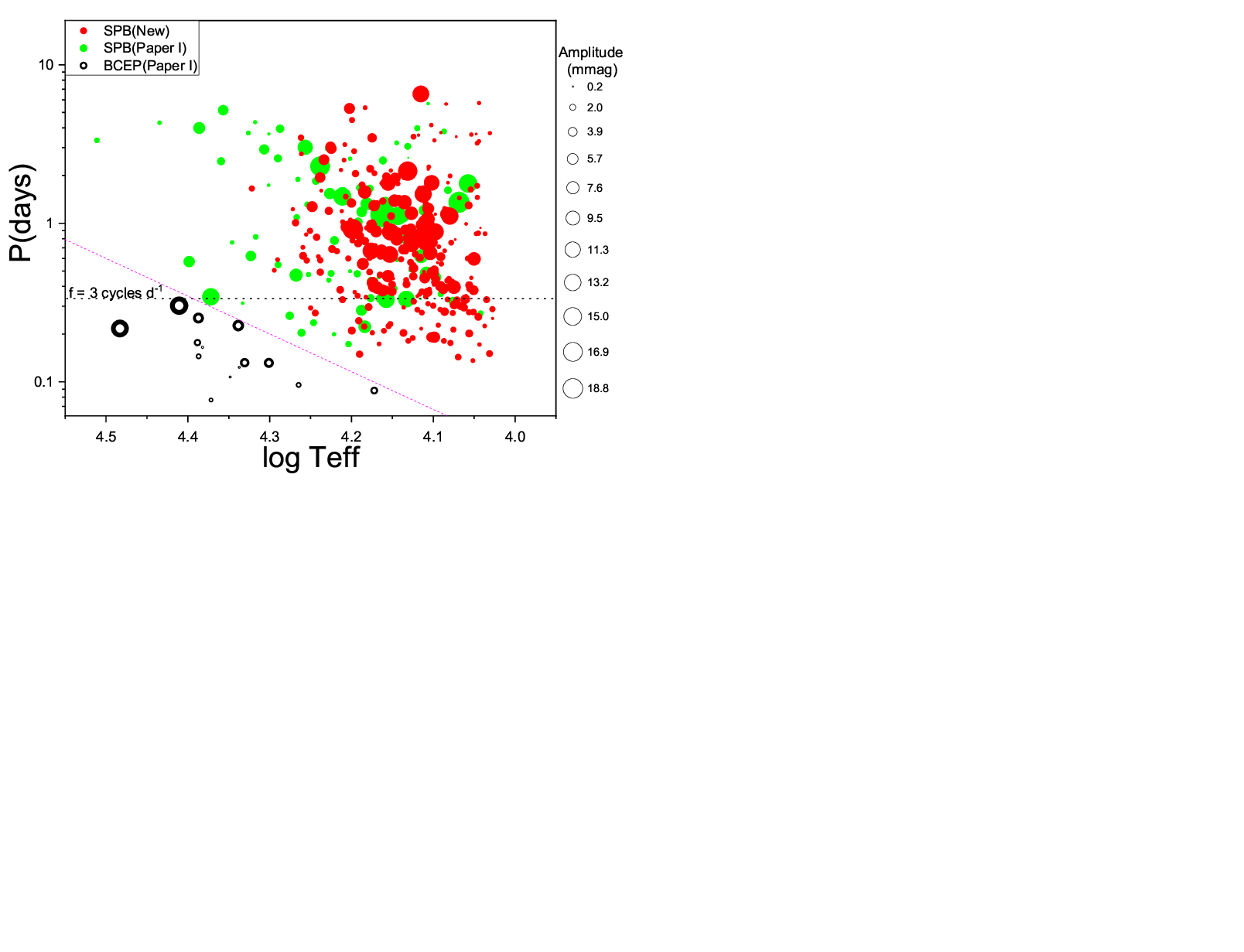}
\caption{The dominant pulsating period and the effective temperature relation diagram of these new SPB stars. Similar symbols to those in Fig \ref{fig:L-T} are used, but the size of the circles denotes their pulsation amplitude of the dominant frequency.}
\label{fig:P-T}
\end{figure*}

\begin{figure*}\centering \vbox to4.0in{\rule{0pt}{5.0in}}
\includegraphics{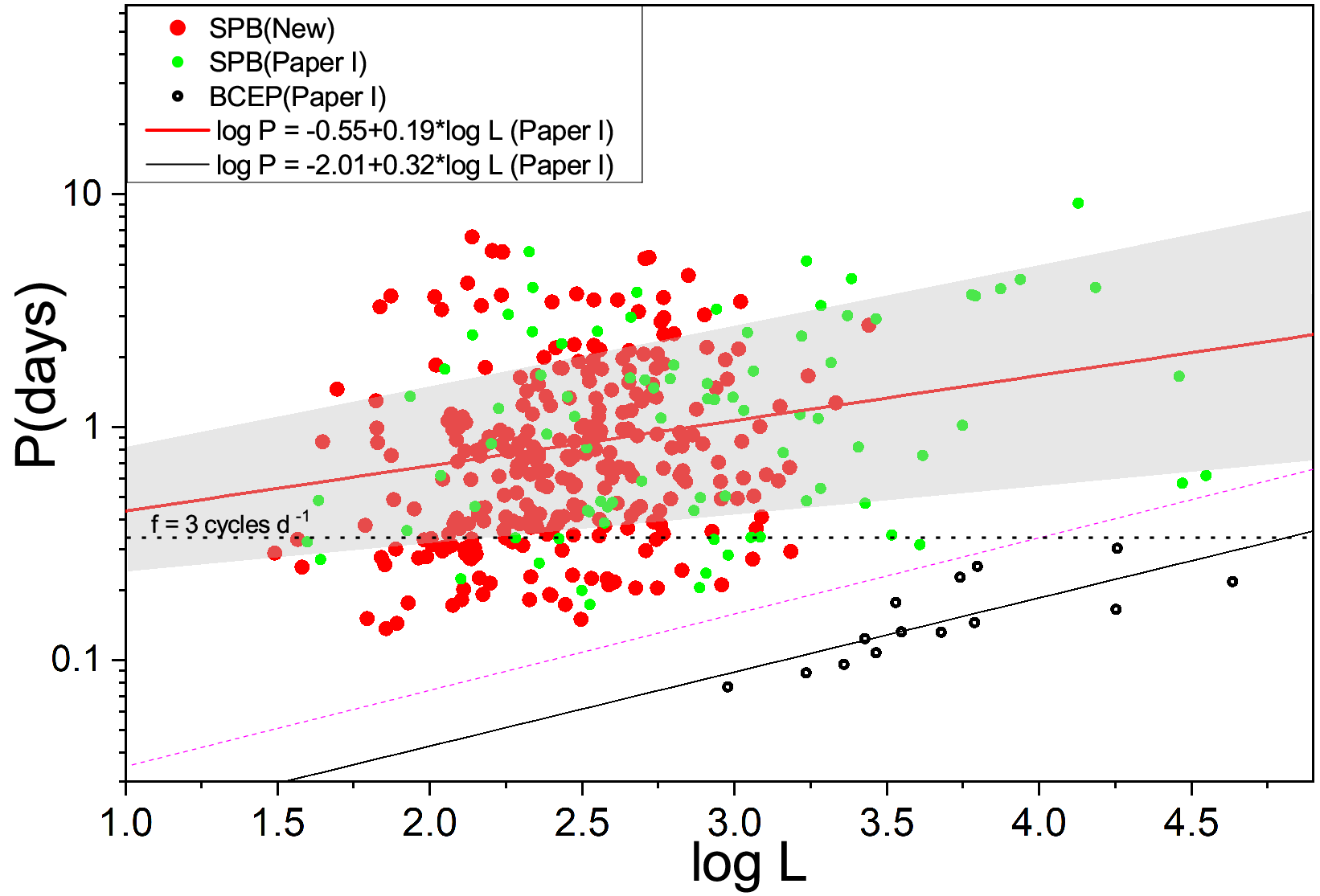}
\caption{The dominant pulsating period and the luminosity relation diagram of these new SPB stars. Symbols are similar to those in Fig \ref{fig:L-T}. The red solid line and the black solid line are the relation derived by Paper I for SPB and BCEP stars, respectively.}
\label{fig:P-L}
\end{figure*}

\section{The properties of these new slowly pulsating B-type stars}

\subsection{The light curves and Fourier spectra}

The Period04 software \citep{2005CoAst.146...53L}, which is based on traditional Fourier analysis, is used to examine the fourier spectra of the light curve for these SPB stars. The signal-to-noise ratio (S/N) of their frequencies is calculated with the residuals at original and a box size of 1 cycles $d^{-1}$, and the standard for frequency detection is S/N $\geq$ 4.6 for a 30-min short cadence \citep{2021AcA....71..113B}. The frequency resolutions are $\delta$f = 1.5/$\Delta$T, where $\Delta$T is the observations time span. Some examples of light curves and Fourier spectra for SPB stars are displayed in Figures \ref{fig:1} and \ref{fig:2}. Columns (8) and (9) of Table \ref{tab:1} list the periods and amplitudes that correlate to the dominant frequency of these SPB stars, and the errors on the last one bits are in parentheses, where the errors are calculated by \citet{1999DSSN...13...28M}.

The suggested variable star type is provided in Column (7) of Tables \ref{tab:1} and \ref{tab:2}. As described in Paper I, there are two outlines of Fourier spectra for SPB stars. One is to only include low-frequencies that are identifiable as g-modes. The other is to have both low- and high-frequencies that can be recognized as g-modes and their combined frequencies \citep{2015MNRAS.450.3015K}, e.i., those high-frequencies are merely harmonics or combinations of low-frequencies. Here the value of 3 cycles $d^{-1}$ is used as a rough criterion to classify high- and low- frequencies.

The rotational modulation is generally due to spots or the stellar wind \citep{2020A&A...639A..81B}, and the respective brightness variations are typically well described with a single sine-function and a couple of its (sub-)harmonics. The eclipse signal in the eclipsing binary is typically much more complex and requires a long series of harmonic terms of the base frequency to describe it, but the variations in the binary star with low orbital inclination (without eclipse) may be similar to rotational modulation stars. However, all of these will not contaminate our sample because those targets with only a single frequency and its harmonic frequencies have been eliminated.

Internal gravity waves (IGWs) are incoherent pulsations excited at the boundary between the convective and radiative zones, and will exhibit stochastic low frequency variability, but coherent pulsation frequencies cannot be detected \citep{2020A&A...639A..81B}.
If a target belongs to one of the following situations, it will be recognized as a candidate for SPB stars and listed in Table \ref{tab:2}: (a) it is difficult to conclude from its limited frequency resolution and S/N that it is g-mode pulsations, rotational modulations, or IGWs, but it is suspected as an SPB star, which will be labeled as 'SPB?'; (b) an SPB star is in a system of eclipsing binary (EB) or rotational variable (ROT), which will be marked as SPB+EB or SPB+ROT.

\subsection{Be stars}

Based on the spectra of LAMOST DR7, we examine the $H_{\alpha}$ emission line profile. A total of 20 objects are identified as Be stars or candidates, and their effective temperature measurement may be impacted by the presence of emission lines. The results are marked as 'Be' or 'Be?' in the parentheses of Column (5) of Table \ref{tab:1} and \ref{tab:2}, which depends on whether the emission line profile is obvious. Fig \ref{fig:Spectra} show example low-resolution spectra of Be star TIC266200717 and candidate TIC270467847 from LAMOST.

There may be some targets with emission characteristics that have not been detected. LAMOST has made multiple observations for some targets, and we found that the $H_{\alpha}$ emission line profile of some targets may not always exist in every observation. This leads to the possibility that some targets with the $H_{\alpha}$ emission line profile cannot be detected by one or two observations. In addition, the emission characteristics of some targets can only be detected in the medium-resolution spectra, but they may only be observed by low-resolution spectra.

\subsection{The luminosity and H-R diagram}

In Column (6) of Tables \ref{tab:1} and \ref{tab:2}, the luminosities of these SPB stars and candidates are displayed, and are computed with the following formulae,
\begin{equation}
log(L_/L_{\odot})=0.4\times{(4.74-M_{V}-BC)}
\end{equation}
\begin{equation}
M_{V}=m_{V}-5\times{log(1000/\pi)}+5-A_{V},
\end{equation}
where the interstellar extinction $A_{V}$ and the parallax $\pi$ are from Gaia DR3, the visual magnitude $m_{V}$ is described in the previous section, and the bolometric correction $BC$ is calculated according to the function of temperature for local thermodynamic equilibrium (LTE) grids given by \citet{2020MNRAS.495.2738P}. The mean errors of $log(L_/L_{\odot})$ are predicted to be around 0.1 dex (e.g., \citet{2020MNRAS.493.5871B}), based on the typical uncertainties 0.05 mas, 0.10 mag, 0.02 mag, and 0.01 mag for $\pi$, $A_{V}$, $BC$, and $m_{V}$, respectively.

Figure \ref{fig:L-T} shows the H-R diagram of these new SPB stars and candidates, Be stars, and those SPB and BCEP stars published in Paper I. Here the stellar evolution code Modules for Experiments in Stellar Astrophysics (MESA, \citet{2011ApJS..192....3P, 2013ApJS..208....4P, 2015ApJS..220...15P, 2018ApJS..234...34P, 2019ApJS..243...10P} is used to produce the evolutionary tracks for the mass of 2.5, 7, and 20 $M_{\odot}$ as well as the theoretical zero-age main sequence (ZAMS) for Z = 0.02. This figure also present the theoretical instability region of SPB and BCEP stars for Z = 0.02 and spherical harmonic degree $l \leq$ 3 from \citet{2007CoAst.151...48M}.

It can be seen in the H-R diagram that these new SPB stars are in the main sequence evolutionary stage with a mass between 2.5 to 7 $M_{\odot}$, a surface effective temperature between 10000 to 21000 K, and a luminosity between 40 to 2850 $L_{\odot}$. Meanwhile, most these new SPB stars are distributed in the instability region of SPB stars, but there are some targets beyond the red edge of theoretical instability region, i.e. these targets in the black dashed box of Figure \ref{fig:L-T}. We compared the light curves and Fourier spectra for these new SPB stars inside and outside the instability region, and found that there are no difference between the two. Figures \ref{fig:1} and \ref{fig:2} show several examples of light curves and Fourier spectra inside and outside the instability region, respectively.

We also examined the projected rotation velocities of those SPB stars to find out what is different about those objects inside and outside the instability region. Since did not obtain a consistent distribution from the three sources, we compared the data from each source separately, as shown in Figure \ref{fig:vsini}. The following two points can be seen in this figure: a) although the distributions from the three sources are different, the number of SPB stars inside the instability region decreases with increasing of projected rotation velocities, and b) regardless of the data source, the number of SPB stars with high projected rotation velocity is significantly over-represented among those outside the instability region.

\subsection{The P-T and P-L diagram}

Due to various reasons, it is necessary to distinguish SPB and BCEP stars from different perspectives. These reasons are, for example, the possible presence of combined high frequencies in SPB stars and independent frequencies in BCEP stars that can be misidentified due to limited frequency resolution, the existence of overlapping region in the instability regions of both, and the potentially inaccurate temperature and luminosity of individual targets.

Figure \ref{fig:P-T} depicts the relation between the dominant pulsation period and the surface effective temperature (P-T) for these new SPB stars and those SPB and BCEP stars published in Paper I. The size of the circles represents the pulsation amplitude of the dominant frequency. This figure shows that these new SPB stars have a period range from 0.14 to 6.5 days and an amplitude range of 0.2-20 mmag in TESS band. They are located in a region separated from these BCEP stars, similar to the SPB stars in Paper I.

Figure \ref{fig:P-L} draws the relation between the dominant pulsation period and the luminosity (P-L) for these new SPB stars, and those SPB and BCEP stars published in Paper I. These new SPB stars are in accordance with the P-L relation derived in Paper I:
\begin{equation}\label{eq:P-L1}
log P = -0.55(\pm0.20)+0.19(\pm0.07)\times{log L}.
\end{equation}

\section{Discussion and conclusion}

Based on the photometric, spectral, and astrometric data of TESS, LAMOST, and Gaia surveys, we detected 286 new SPB stars and 21 candidates, 20 of which are Be stars or candidates. We then analyzed their Fourier spectra and obtained their dominant pulsation periods and amplitudes, and also estimated their luminosities from their parallaxes and visual magnitudes. It is shown that these SPB stars have luminosities between 40 and 2850 $L_{\odot}$ and effective temperatures ranging from 10000K to 21000K. Their pulsation periods are from 0.14 to 6.5 days with amplitude ranges of 0.2-20 mmag in TESS band.

In the H-R diagram, most of these SPB stars are in the main sequence evolutionary stage and with mass rangs of 2.5-7 $M_{\odot}$. The H-R diagram positions of these pulsators also indicate that most of them are distributed in the instability region of SPB stars, but there are some targets beyond the red edge of the theoretical instability region. We compared the light curves and Fourier spectra for these new SPB stars inside and outside the instability region, and found that there is no difference between these two. However, when we examined the distribution of projected rotation velocities, we found that the number of SPB stars with high projected rotation velocity is significantly over-represented among those outside the instability region. It implies that those stars outside the instability region should be fast-rotating SPB stars whose measured effective temperatures are reduced by gravity darkening at their equators \citep{1924MNRAS..84..665V}. This conclusion is consistent with previous research, for example, \citet{2014A&A...569A..18S, 2022MNRAS.515..828S, 2005MNRAS.360..465T}.


In addition to the H-R diagram, the P-T and P-L diagrams can be crucial for distinguishing SPB stars from BCEP stars (Paper I). It is indicated that these targets follow the distribution of the SPB stars in Figures \ref{fig:P-T} and \ref{fig:P-L}. The P-L relation determined in Paper I is further supported by the fact that these new SPB stars completely conform to this relation. However, the targets of both this paper and Paper I are distributed in a wider band on the P-L diagram, which means that SPB stars may contain different subclasses, similar to RR Lyrae stars containing RRa, RRb, RRc, and RRd.

The discovery of these new SPB stars increases their total number by over 60\%, which are significant samples for further investigating the structure and evolution of intermediate-mass and even massive stars by asteroseismology. It will be required to investigate these objects in greater detail and from various views in the future.

\acknowledgments
This work is partly supported by Chinese Natural Science Foundation (Nos. 11933008, 12103084 and 12273103) and the basic research project of Yunnan Province (Grant No. 202201AT070092).
The spectral data of this paper were observed by the Large Sky Area Multi-Object Fiber Spectroscopic Telescope (LAMOST).
This work has made use of data from the European Space Agency (ESA) mission Gaia (https://www.cosmos.esa.int/gaia), processed by the Gaia Data Processing and Analysis Consortium (DPAC, https://www.cosmos.esa.int/web/gaia/dpac/consortium). Funding for the DPAC has been provided by national institutions, in particular the institutions
participating in the Gaia Multilateral Agreement.
The TESS data presented in this paper were obtained from the Mikulski Archive for Space Telescopes (MAST) at the Space Telescope Science Institute (STScI). STScI is operated
by the Association of Universities for Research in Astronomy, Inc. Support to MAST for these data is provided by the NASA Office of Space Science. Funding for the TESS mission is provided by the NASA Explorer Program.
We thank the anonymous referee for valuable suggestions and comments that significantly improved the paper.


\end{document}